# Direct observation of domain wall surface tension by deflating or inflating a magnetic bubble

Xueying Zhang[1,2], Nicolas Vernier[2], Weisheng Zhao[1,*], Haiming Yu[1], Laurent Vila[3] and Dafiné Ravelosona[2]

[1] Fert Beijing Institute, BDBC, School of Electronic and Information Engineering, Beihang University, Beijing, China

[2] Centre for Nanoscience and Nanotechnology, University Paris-Saclay, 91405 Orsay, France

[3] Univ. Grenoble Alpes, CEA, CNRS, Grenoble INP, INAC, SPINTEC, France

*weisheng.zhao@buaa.edu.cn

**Abstract**

The surface energy of a magnetic Domain Wall (DW) strongly affects its static and dynamic behaviours. However, this effect was seldom directly observed and many related phenomena have not been well understood. Moreover, a reliable method to quantify the DW surface energy is still missing. Here, we report a series of experiments in which the DW surface energy becomes a dominant parameter. We observed that a semicircular magnetic domain bubble could spontaneously collapse under the Laplace pressure induced by DW surface energy. We further demonstrated that the surface energy could lead to a geometrically induced pinning when the DW propagates in a Hall cross or from a nanowire into a nucleation pad. Based on these observations, we developed two methods to quantify the DW surface energy, which could be very helpful to estimate intrinsic parameters such as Dzyaloshinskii-Moriya Interactions (DMI) or exchange stiffness in magnetic ultra-thin films.

Magnetic Domain Walls (DWs) in magnetic ultra-thin films have attracted a lot of interest due to its perspective to develop high-density non-volatile memory and logic applications[1–3]. A DW is the interface separating two magnetic domains, which can be moved using a magnetic field or a spin polarized current. The fundamental behaviour of a DW is similar to that of many other type of interfaces in physics, such as vortices in superconductors, soap films or surfaces of liquids[4,5]. The properties of these interfaces have been intensively studied [6–12]. The main parameter to explain the observed behaviour is the surface energy γ. Whereas, in magnetism, the effect of magnetic DW surface tension (this tension is the force associated to the surface energy) is difficult to be directly observed, although it plays a very important role. For instance, the motion of DWs in the so-called creep regime results from the competition between surface energy γ, the pinning energy and the Zeeman energy from the applied magnetic field. However, in the universal law describing DW velocity, γ is hidden in phenomenological constants[13,14]. In addition, some phenomena related to DW surface energy have not been well understood, such as DW pinning in an artificial constriction[15–17]. In particular, the surface tension of DW plays a critical role in the topological transition of domain structures, for example, the transition from domain stripes to skyrmionic bubbles[18]. Moreover, the stabilization of these skyrmionic bubbles is directly determined by the competition between the DW surface tension and the dipolar interaction[19].

In thin films with perpendicular magnetic anisotropy, the surface energy of a DW is given by $\gamma = 4\sqrt{AK_{eff}}$, where A is the exchange stiffness and $K_{eff}$ is the effective anisotropy energy, assuming that the DW is of Bloch type[20]. Recently, Dzyaloshinskii-Moriya Interactions (DMI) has been intensively studied since it is found to be essential in the formation of stable skyrmions in magnetic thin films[18,21] or to obtain the high velocity of the DW motion driven by the spin Hall current[22,23]. This interaction results in an additional term in the expression of the surface energy γ that is proportional to the DMI coefficient[21,24,25]. It becomes crucial to measure γ in order to better understand the role of different energy terms involved in the physics of DWs and directly obtain intrinsic magnetic parameters such as the exchange stiffness A or the DMI coefficient. However, a precise and direct measurement of the surface energy is missing. In the past, the method to measure γ usually relies on gauging the size of magnetic domains at a demagnetization state[26,27,28]. Nevertheless, the structure of domains strongly depends on the underlying pinning potential and the way the demagnetizing state is reached, which makes this approach unreliable for γ mesurement.

In this work, we show that a magnetic bubble can spontaneously collapse in zero fields due to the DW surface tension and can be stabilized using an external field. The interaction between two bubbles is also investigated. DWs depinning mechanism at the Hall cross or at the entrance of the nucleation pad is studied and explained in terms of the DW surface energy. Based on these observations, two approaches have been proposed to measure the DW surface energy, one is through the dependence of the stabilizing field on the bubble size and the other is through the dependence of the depinning field on the nanowire width.

## RESULTS

### Spontaneous contraction of the magnetic bubble

The sample studied is a Ta(5nm)/CoFeB(1nm)/MgO (2nm)/Ta(5nm) multilayers stack with perpendicular anisotropy. It was patterned into a square connected with a narrow wire, as shown in Fig. 1a. The size of the magnetic square is 20 μm ×20 μm and the width of the wire in different structures varies from 200 nm to 1.5 μm. After obtaining a DW in the wire, we injected the DW from the narrow wire into the square by using a large field pulse. After the pulse, a semi-circular bubble domain was obtained, as shown in Fig. 1b.

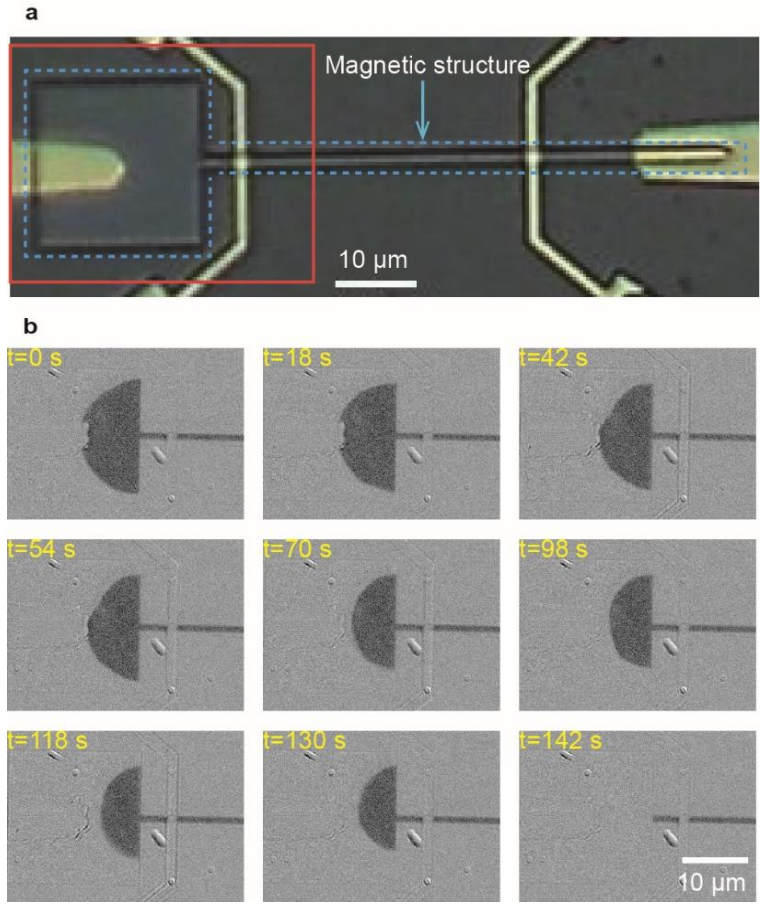

**Figure 1 | Structure of the sample and the evolution of a semicircular domain bubble. a,** Optical image of one of the samples studied. The width of the wire for this sample is 1 μm. The zone surrounded by the blue dashed line is the magnetic structure and that surrounded by the red line corresponds to the area viewed in (b); **b,** Kerr images to show the spontaneous contraction of the semicircular domain bubble in a zero external field. A movie showing this dynamic contraction can be downloaded online (see supplementary information).

Subsequently, although the magnetic field was zero, we found that the bubble could contract spontaneously, until the DW had returned to the entrance of the nucleation pad, i.e. the neck. Furthermore, the speed of contraction depended on the radius of curvature of the DW circle. The smaller the radius, the faster was the contraction. As shown in Fig.1b, when the radius of the bubble is approximately 9 μm, the DW contraction is observable only after 20 s. However, when the radius of the DW bubble shrinks to 4 μm, the DW returns to the neck after several seconds. After returning to the neck, the DW does not move anymore.

We checked that there was no remnant parasitic field that could induce such an effect. The sample holder was made of a non-magnetic material, and a Hall probe was used to check the magnetic field around the sample. In addition, the experiment was conducted with both directions of the magnetization of the bubble and the result was the same, which would not have been the case if there had been a remnant field.

A DW can be seen as an elastic membrane with energy γ per unit area[29]. According to the physics of membranes, assuming an isotropic pressure, equilibrium of a membrane is obtained when the difference of pressure between both sides is equal to γ/R, where R is the curvature of the membrane. Consequently, here, the interface between the domain bubble and the adjacent domain is a cylindrical surface of radius R, as shown in Fig. 2. To begin with, let us remember that for a magnetic system, the effect of magnetic field H on a DW can be seen as a pressure of magnitude $P_H = 2\mu_0 H M_S$.

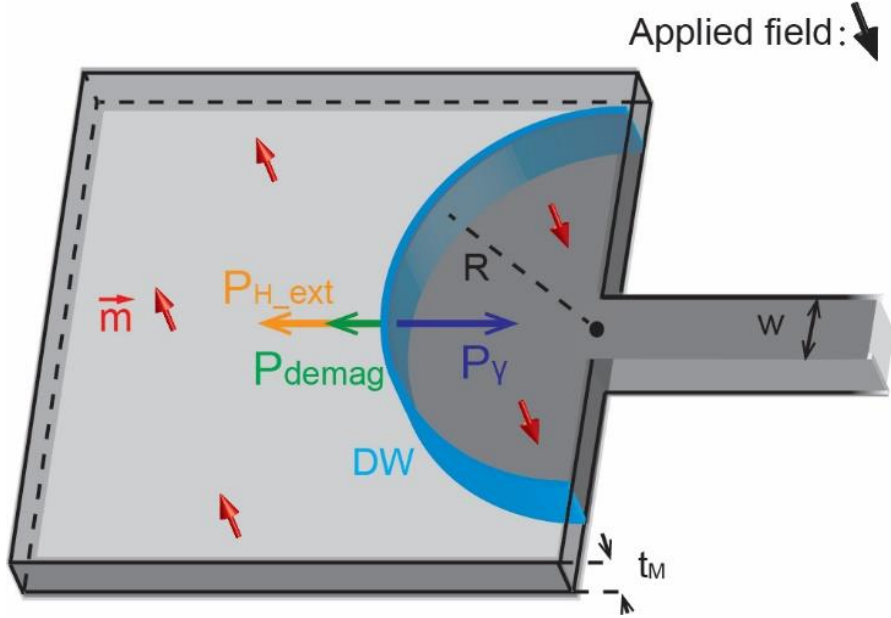

**Figure 2 | 3D sketch of the profile of the semicircular magnetic bubble stabilized in the square.** Arrows on the DW indicate the forces induced by the different pressures on this part. Since the pressures are isotropic, the direction of the local forces points perpendicularly to the surface of the DW and changes according to the position on the DW.

Now, according to the Laplace-Young equation[4,5,29], when the DW presents a curvature, a pressure is induced on the DW due to its surface energy. In the case of a cylindrical surface, this pressure is given by:

$$P_\gamma = \gamma/R \qquad (1)$$

Here, this pressure is high enough to induce some movement of the bubble in the creep regime. Note that it remains a relatively low force; the movement is possible because CoFeB is very soft[30]. Moreover, the bubbles created had a quite small radius (<10 μm), which induces a quite large Laplace pressure. In agreement with $P_\gamma = \gamma/R$, the spontaneous contraction was faster as the radius of the domain reduced.

## Stabilization of the magnetic bubble and estimation of the DW surface energy

To quantify the Laplace pressure, the external field needed to compensate for this pressure and stabilize domain bubble was measured as follows: First, a very small static field was applied. Second, we created a semicircular bubble domain using a large field pulse. In the experiments, the initial size of the semicircular domain bubbles could be controlled through the magnitude and the duration of the field pulse when blowing the DW into the square. To check whether it was stable or not under the static field, we waited for 60s. If no movement was detected after 60s (for example, in the case of Fig. 3b, there was a movement), the state could be considered stable. Then the static field was increased (or decreased) in a steps of 20 μT, until a movement was detected. In Fig. 3c, the critical fields for expansion and for contraction have been plotted as a function of the inverse of the radius of the domain bubble. The equilibrium field is between the two clouds of critical fields for expansion and for contraction. As could be expected, this field had to be applied along the magnetization direction inside the bubble.

It was found that the magnitude of the external field required to stabilize the semicircular domain bubble depended on the size of the bubble. As the radius of the domain bubble scaled down, the stabilization field increased sharply.

Theoretically, a circular DW is at equilibrium only when the pressure due to the magnetic field exactly cancels out the excess pressure from the DW tension. Since the elastic force increases linearly as the inverse of the radius of the DW bubble, the field required to stabilize the semi bubble should be inversely proportional to the radius. In Fig. 3c, the stabilizing field appears as the boundary between the expansion points and contraction points. We can see that there is a very good agreement between the predicted behaviour and the experimental one, the slope being $k_{equ} \approx 1.2\ mT \cdot \mu m$.

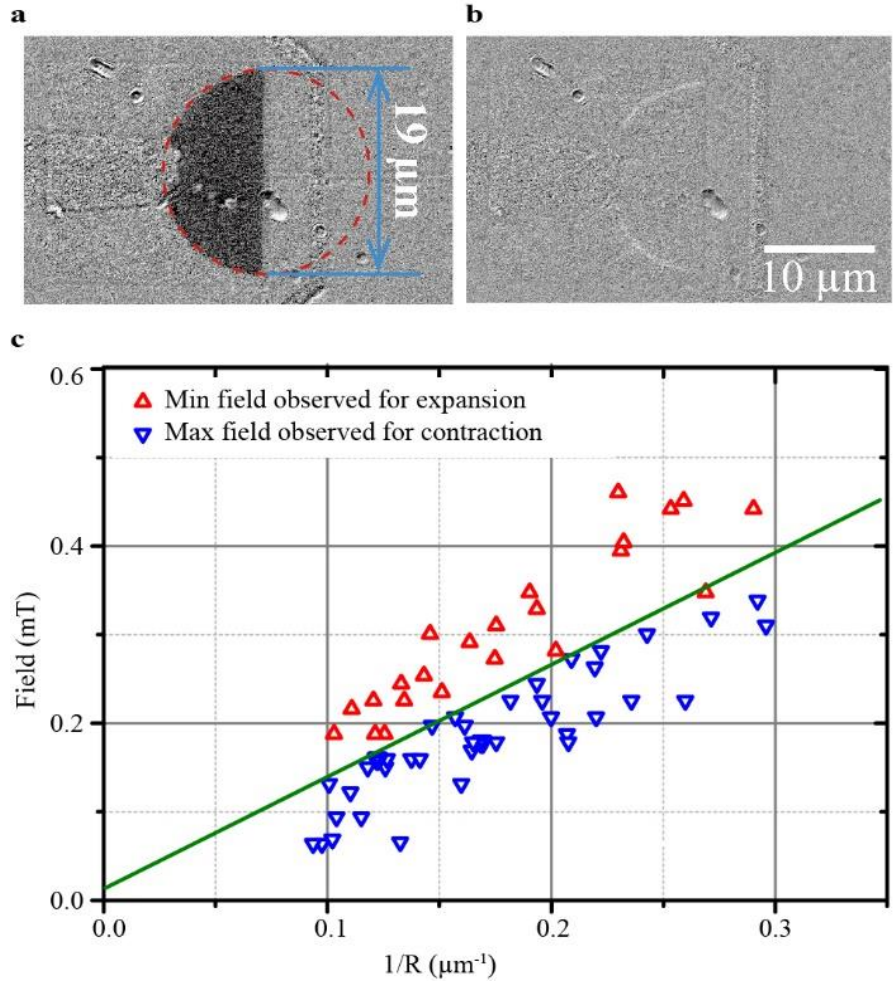

**Figure 3 | Stabilization of the domain bubble with an external field. a,** A semicircular bubble is stabilized by an external field $\mu_0 H_{ext} = 0.11\ mT$ and no DW motion is observed during 60s; **b,** The external field decreases to $\mu_0 H_{ext} = 0.094\ mT$ T and the domain bubble contracts slightly. This image is the difference between two images: the first one was acquired before reducing the applied field and the second one was acquired 66s after reducing the applied field. The white circular trace shows the DW displacement; **c,** Critical field for expansion and contraction as a function of the inverse of the radius of the semicircular domain bubble. Two structures associated with a 600 nm wide wire were used to obtain more statistics.

However, one should be careful because there are two contributions for the magnetic field: the externally applied field and the demagnetizing field. We have numerically calculated the demagnetizing field using the concept of magnetization current (see supplementary information) and found that the demagnetizing field can be approximated by a linear law $\mu_0 H_{demag} \approx k_{demag} \frac{1}{R}$. For a radius R from 3 μm to 10 μm, the slope $k_{demag} = 1.24\ mT \cdot \mu m$.

From the above considerations, the equilibrium of the pressures can be written as follows:

$$2\mu_0 H_{ext} M_S + 2\mu_0 H_{demag} M_S - \frac{\gamma}{R} = 0 \qquad (2)$$

Using the linear laws found for $\mu_0 H_{demag} = k_{demag}/R$ and the equilibrium field $\mu_0 H_{equ} = k_{equ}/R$ and simplifying by 1/R, the surface energy of the DW is given by:

$$\gamma = 2M_S(k_{equ} + k_{demag}) \qquad (3)$$

From equation (2), we get γ ≈ 5.4 mJ/$m^2$. From[31,32], the exchange stiffness A in this type of material was found to be between 10 and 28 pJ/m. For our sample, the effective anisotropy was measured as $K_{eff} = 2.2 \times 10^5 J/m^3$ [33]. According to the formula $\gamma = 4\sqrt{AK_{eff}}$, it gives γ between 6 and 10 mJ/m2. It can be seen that our value is in good agreement with the theoretically calculated one. Note that the formula used here to calculate the surface energy γ applies only to films without DMI. In fact, the DMI constant measured in our system is less than 0.01mJ/$m^2$, meaning that its effect on the DW surface energy is negligible[34]. This result is also confirmed by Brillouin light scattering experiments (not yet published).

## Interaction of two bubbles

As suggested before, the behaviour of magnetic bubbles induced by the surface energy is similar to that of soap bubbles. There is a well-known experiment in which two soap bubbles are connected to each other; the result is not what one might expect intuitively. Instead of obtaining two bubbles of equal size, the smaller bubble empties itself into the bigger bubble and disappears[4,5,35].

Here, we observed the same phenomenon. We simultaneously created two semicircular bubbles with different sizes in a magnetic square, as shown in Fig. 4. An external field $\mu_0 H_{ext} = 0.41 mT$ was applied to avoid the collapse of both bubbles. In the beginning, the smaller magnetic bubble was stable while the larger one expanded slowly. Once the two bubbles were close enough, there was a repulsive pressure between the two bubbles due to dipolar interaction[36]. Similar to the two connected soap bubbles, the bigger bubbles expanded and the smaller one collapsed and disappeared.

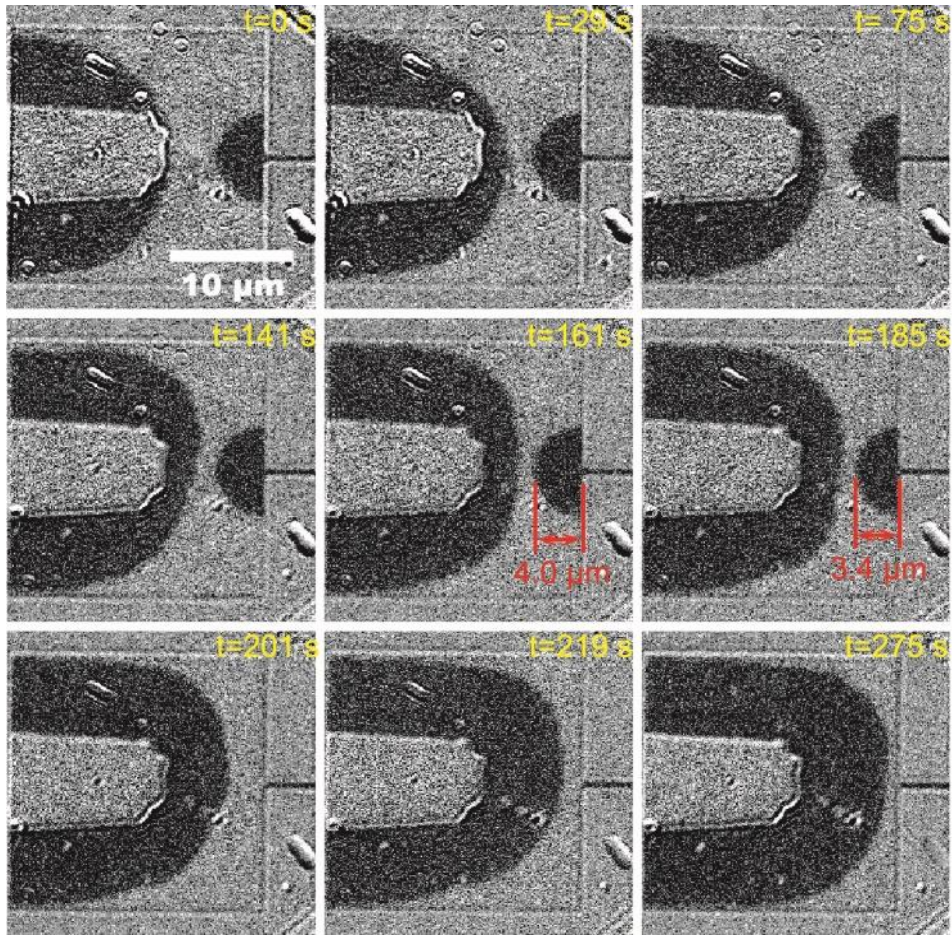

**Figure 4 | Interaction of two semicircular domain bubbles.** Under an applied field 0.41 mT, the bigger bubble expands and squeezes the smaller one because of dipolar interaction. The movie on this dynamic process can be downloaded online (see supplementary information).

It can be explained by the pressure induced by the DW surface energy. This pressure tends to reduce the sizes of the bubbles. This pressure is higher for the smaller bubble, as predicted by equation (1). Owing to the repulsion between the two bubbles, we cannot have both bubbles expanding together; one of them must contract. It is the smaller one that contracts because of the higher pressure due to surface energy.

## DW depinning at the neck or Hall crosses

Now, we are coming to what may be the most interesting result: DW injection into the nucleation pad or DW depinning at a Hall cross. Again, the surface energy appears to be a fundamental effect involved in these processes. Note that the Hall cross is a very commonly used structure in magnetic experiments[37–39].

Indeed, when injecting a DW from the narrow wire into the square, we found that the DW was pinned at the neck and would not be depinned until the applied field reached a threshold value. As shown in Fig. 5a the DW was moved by a field pulse. After the pulse, we could see that the DW stopped at the neck. We tried to send more pulses of the same magnitude, but no more movement occurred; the DW remained pinned at the neck. To depin the DW, we had to increase the magnitude of the field; the critical field required to achieve the movement is called the depinning field $H_{dep}$ here. It should be noted that the duration of the field pulse was always 5 µs in these measurements. We repeated this experiment several times for each sample. The depinning field was almost always the same. For Hall crosses, the behaviour was the same (Fig. 5b): pinning, if the magnetic field was below a critical field, and movement above it.

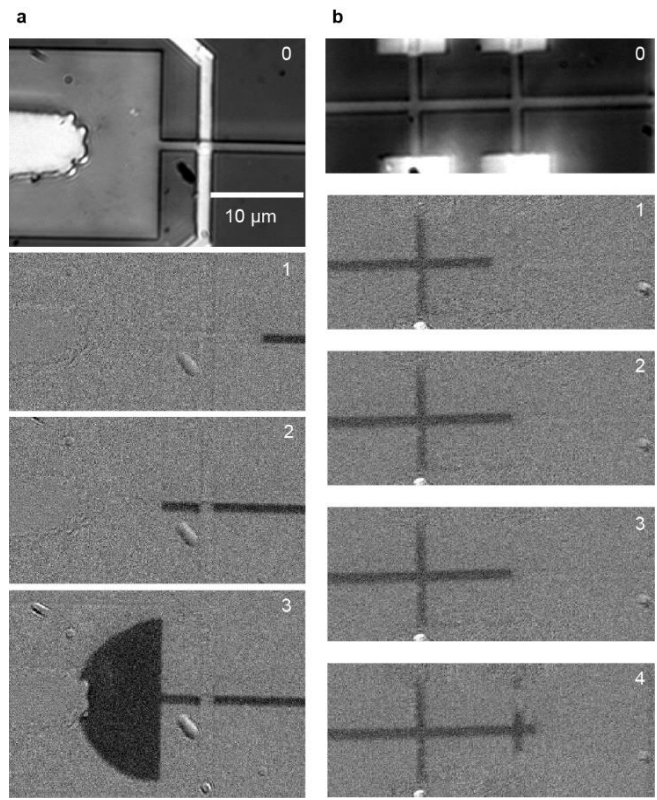

**Figure 5 | DW depinning at the neck (a) and at a Hall cross (b). a,** Image n°0 is the raw image of a structure and is used as the reference image to remove the background and obtain the Kerr image. Image n°1 shows the initial position of the DW. After a field pulse of 5.8mT and 5 µs, the DW moves to the neck (n°2) and trapped there. After increasing the magnitude of the pulse to 5.9 mT, the DW is depinned and expands to a semicircular bubble (n°3); **b,** Image n°0 is the raw reference image of a 1 µm wide wire. Between each picture from n°1 to n°3, a pulse of 5.9 mT and 5 µs is applied and the DW is moved in the wire but pinned at the Hall cross. The DW is depinned with a field pulse of 6.0 mT and 5 µs (n°4).

The interesting point is that the depinning field depends on the width w of the wire. We have plotted $H_{dep}$ in Fig. 6a. It can be seen that there is a linear dependence of $H_{dep}$ as a function of the inverse of the width. Furthermore, the dependence is almost the same for the Hall cross or for the neck of the pad. Let us note that when 1/R tends to zero, the depinning field tends to $\mu_0 H_0 = 3\ mT$. This is explained by the velocity law. Below 3mT, for pulses of 5μs, no effective movement is possible. Movement requires a field larger than a minimum value $H_P$, which can be estimated at 3 mT [33]. Indeed, it is true that even for magnetic fields lower than 3mT, a DW can move in this kind of sample. But, after the end of the pulse, since it can not go very far with a 5 μs pulse, it is driven back by the surface energy, as described previously in this paper. Typically, for the Hall cross, with a wire width between 200 nm and 1.5 μm, the DW has to propagate over at least 200 nm (up to 1.5 μm) to reach the opposite wire, which means that a velocity greater than 40mm/s (300mm/s) and a field greater than 3mT (almost the same value) are required.

Now, our experiments show that the pinning is due to surface energy. Indeed, we can do the same analysis as the stabilization of a domain bubble; in this case, the minimum radius of the bubble is w/2, where w is the wire width (see Fig. 6b&c). Therefore, the maximum pressure induced by the Laplace interface force is 2γ/w. Movement and depinning from the neck can occur only if the overall pressure due to the magnetic field and Laplace pressure is larger than the pressure induced by $H_0$ alone.

$$2\mu_0 H_{ext} M_S + 2\mu_0 H_{demag} M_S - \frac{2\gamma}{w} \geq 2\mu_0 H_0 M_S \qquad (4)$$

For the magnetic field, there are two contributions: the externally applied field and the demagnetizing field. From this equation, the depinning field is predicted to be:

$$H_{dep} = \frac{\gamma}{\mu_0 M_S w} - H_{demag} + H_0 \qquad (5)$$

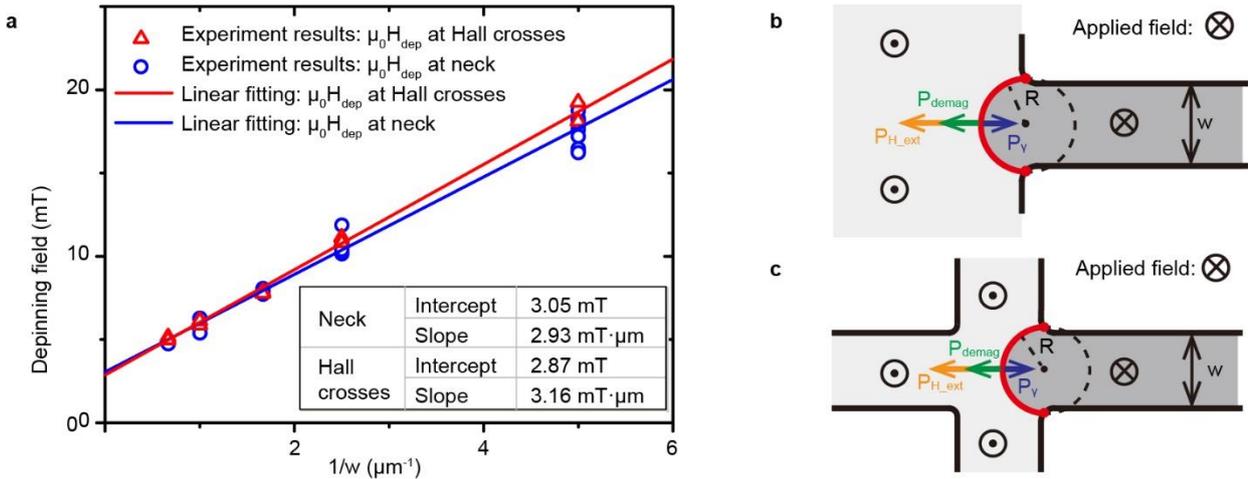

**Figure 6 | Experimentally measured DW depinning field (a) and pressures acting on the DW pinned DW at the neck (b) and at a Hall cross (c). a,** DW depinning field as a function of the inverse of the wire width. Blue: DW depinning at the neck, red: DW depinning at the Hall crosses; **b & c,** sketches of pressures acting on the DW pinned at the corners of a neck or a Hall cross, respectively. It can be seen that the minimum radius (corresponding to the highest Laplace pressure possible) of these arcs of circle is w/2.

As before, H$_{demag}$ can be numerically calculated and we have found again an approximate linear law: $\mu_0 H_{demag} \approx k_{demag}/w$, where $k_{demag} = 0.84 \, mT.\mu m$ in the case of pinning at the neck and $k_{demag} = 0.58 \, mT.\mu m$ at the Hall cross.

$$H_{dep} = \left(\frac{\gamma}{\mu_0 M_S} - k_{demag}\right)\frac{1}{w} + H_0 \quad (6)$$

This prediction fits perfectly with our experimental results. The last thing to check is the value of γ. We have found an experimental slope k=2.9 mT·μm for necks (3.2 mT·μm for Hall crosses). The surface tension γ can be calculated by identifying the slope of equation (6) to experimental k:

$$\gamma = \mu_0 M_S (k + k_{demag}) \quad (7)$$

It gives $\gamma = 4.1 \, mJ/m^2$, which is in good agreement with our previous results. The slight discrepancy could arise from the edges of the Hall crosses, which are probably slightly rounded, inducing a small enlargement of the width of the wire at the entrance of the cross or nucleation pad.

Notice that the experiments we have presented here present a new and very interesting method of measuring the surface energy of DWs. One needs to prepare some pads or Hall crosses by lithography with wires of different widths. Then, by checking the depinning field as a function of the width, the surface energy of the DW can be deduced.

## Discussion

The surface energy of the domain walls can become a very important parameter in some cases. First, we have shown that a semi-circular bubble in a small square becomes unstable and disappears spontaneously in zero external fields in soft magnetic films, due to the Laplace pressure induced by the surface energy. Checking the magnetic field required to stabilize the semicircular domain, we have been able to estimate the surface energy of the DW, which was found to be 5.4 mJ/m². Second, we have observed the effect of interaction between two semicircular bubbles in a small square when an external field is applied. The bigger bubble grows continuously. When the two bubbles are close enough, the dipolar repulsion induces the contraction of the smaller one until it has disappeared. It is again due to the Laplace pressure, which is bigger for the smaller bubble. To finish, we have identified the origin of the pinning of the DW at the Hall crosses or at the neck between a narrow wire and a large pad. The Laplace pressure induced by the surface energy has to be overcome by the externally applied field. The depinning field scales as 1/w, where w is the width of the wire. Plotting the depinning field as a function of 1/w appears a very good method to measure the DW surface energy. Through this method, we found the surface energy to be 4.1 mJ/m², which agrees with our first value. Our experiments show the importance of the DW surface energy: it explains the pinning on nanowires due to defects or notches, especially at Hall crosses, which is critical for applications such as racetrack memory. Moreover, our method for the surface energy measurement could be very helpful in the study of thin films presenting DMI, since these interactions change the surface energy drastically and could enable faster DW motion[40].

## Methode

The sample studied is a Ta(5nm)/CoFeB(1nm)/MgO (2nm)/Ta(5nm) multilayers stack with perpendicular anisotropy. It was annealed at 300 °C for two hours and patterned into a square connected with a narrow wire, as shown in Fig. 1a. The size of the magnetic square is 20 μm ×20 μm and the width of the wire in different structures varies from 200 nm to 1.5 μm. Several properties of the samples were experimentally characterized: saturation magnetization $M_S = 1.1 \times 10^6 A/m$, effective anisotropy $K_{eff} = 2.2 \times 10^5 J/m^3$ and the DW width was estimated to be Δ= 10.7 nm [33]. Several golden electrodes were added on the magnetic structure for the electrical tests.

A Kerr microscope was used to view the DW motion. Its resolution was 500nm, and it was able to locate a DW with this precision on nanowires as narrow as 200nm, as no other magnetic material was around the nanowire.

We used two mini coils with diameter about 6mm to generate the magnetic field: one could reach 130mT with a rise time of 2 μs, making it possible to have magnetic pulses shorter than 10 μs, which is required to avoid the reversal the whole nanowire after DW nucleation. The second coil had a rise time of 0.23 μs, and could reach a field of 25 mT. With this second coil, using pulses as short as 1 μs, we were able to move the DW at any position we wanted after nucleation.

The sample was placed in a completely insulating holder to avoid any eddy current screening the magnetic field when the magnetic pulses were generated.

## ACKNOWLEDGEMENTS

This work was supported in part by the National Natural Science Foundation of China (61501013, 61571023 and 61627813) and the China Scholarship Council. The authors also gratefully acknowledge the International Collaboration Projects 2015DFE12880 and B16001. We also acknowledge the support from the Renatec network for the sample fabrication.

**Additional information:** **S**upplementary information and movies can be downloaded online.
**Competing financial interests:** The authors declare no competing financial interests.

# Supplementary information

## Calculation of the demagnetizing field

We have numerically calculated the demagnetizing field using the concept of magnetization current[1,2]. As shown in Fig. SI-1a, magnetic domains can be infinitely divided into small magnetic elements. Each element is equivalent to a ring current element, with current $I = t_M M_S$, where $t_M$ is the thickness of magnetic layer and $M_S$ is the saturated magnetization. For a uniformly magnetized out-of-plane domain, the current of one magnetic element can always be cancelled out by the current of the neighboring ones, except for elements in the edge of domains. Therefore, the magnetization currents are zero inside or outside the domain. The only non-zero contributions are on the edge of the domain, where there is a line of current parallel to the edge, $I = t_M M_S$. The demagnetizing field is equal to the Oersted field produced by this current. For a semicircular magnetic bubble with radius R in a magnetic square, the electrical circuit is plotted in Fig.SI-1b. Noting that the DW width in our sample is estimated to be 10.7 nm (ref.3). This width is considered as the distance between two domains with opposite magnetization, i.e. the distance between the two circular current circuits.

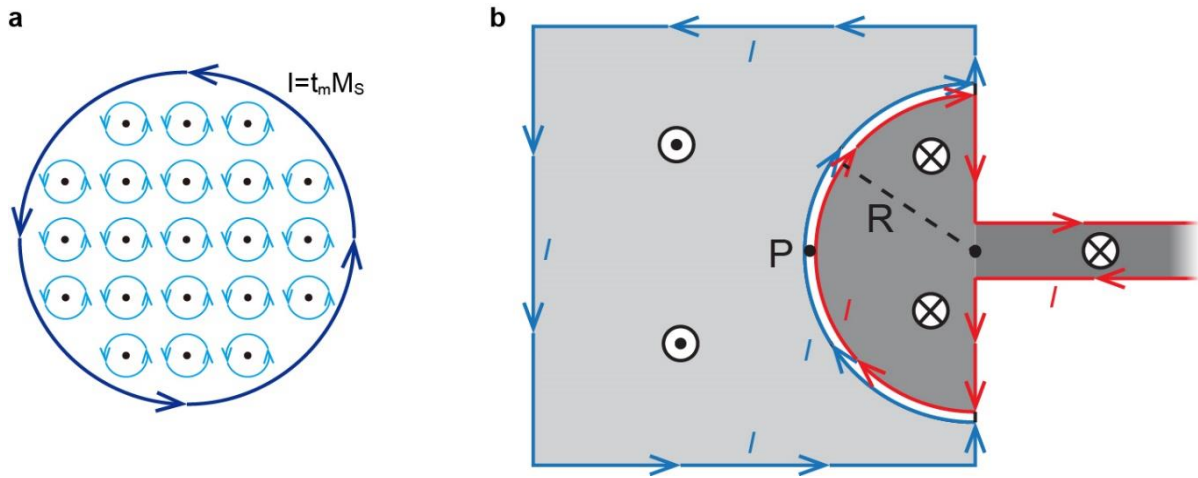

**Figure. SI-1. Magnetization current of the magnetic structure. a,** Sketch to show the concept of magnetization current. **b,** Electrical circuit identical to the current created by the magnetization at the edges of the structure and along the DW. Then, the stray field can be calculated using the Biot-Savart law.

Using MATLAB, we calculated the demagnetizing field in the top of the semi bubble (point P shown in Fig SI-1b) with R varying from 3 μm to 10 μm. The result is plotted in Fig. SI-2. It can be seen that, in this range, this demagnetizing field can be approximated by a linear law, of slope $k_{demag} \approx 1.24\ mT \cdot \mu m$.

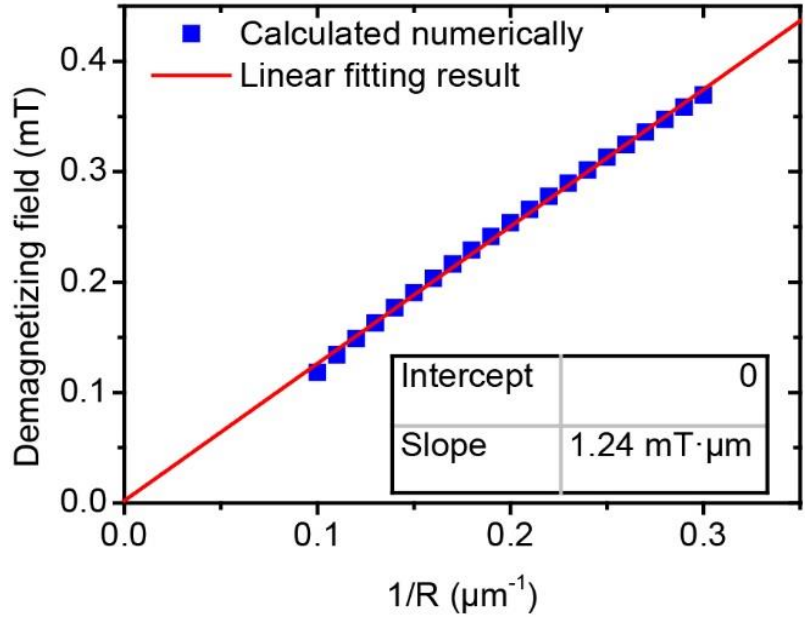

**Figure. SI-2. Numerically calculated demagnetizing field at point P as a function of the inverse of bubble radius.**

For the cases of the DW pinning at the neck or at a Hall cross, we estimated the demagnetizing field using the same method and found a similar results. In the critical condition for the DW depinning, radius of DW equals to a half of the wire width, i.e., R=w/2. The demagnetizing field are approximately linear to the inverse of wire with, $\mu_0 H_{demag} \approx k_{demag}/w$, where $k_{demag} = 0.84\ mT.\mu m$ in the case of pinning at the neck and $k_{demag} = 0.58\ mT.\mu m$ at the Hall cross.